\documentclass[twocolumn,showpacs,preprintnumbers,amsmath,amssymb]{revtex4}
\usepackage{graphicx}% Include figure files
\usepackage{dcolumn}% Align table columns on decimal point
\usepackage{bm}% bold math

\begin{document}

%\preprint{hep-th/yymmnnn}

\title{The generalized uncertainty principle in the
presence of extra dimensions}% Force line breaks with \\
\author{Benrong Mu, Houwen Wu}
\author{Haitang Yang}%
 \altaffiliation[Also at ]{China Institute for Advanced Study,
Central University of Finance and Economics, Beijing, 100081, China.}%Lines break automatically or can be forced with \\
 \email{mubenrong@uestc.edu.cn,  iverwu@uestc.edu.cn,  hyanga@uestc.edu.cn}
\affiliation{%
School of Physical Electronics, University of Electronic Science and
Technology of China,  Chengdu,
610054, China\\
}%

%\date{\today}% It is always \today, today,
             %  but any date may be explicitly specified

\begin{abstract}
We argue that in the Generalized Uncertainty Principle (GUP) model,
the parameter $\beta_0$ whose square root, multiplied by Planck
length $\ell_p$, approximates the minimum measurable distance,
varies with energy scales. Since minimal measurable length and extra
dimensions are both suggested by quantum gravity theories, we
investigate models based on GUP  and one extra dimension,
compactified with radius $\rho$. We obtain an inspiring relation
$\sqrt{\beta_0} \ell_p/\rho \sim {\cal O}(1)$. This relation is also
consistent with predictions at Planck scale and usual quantum
mechanics scale. We also make estimations on the application range
of the GUP model. It turns out that the minimum measurable length is
exactly the compactification radius of the extra dimension.
\end{abstract}

%\pacs{Valid PACS appear here}% PACS, the Physics and Astronomy
                             % Classification Scheme.
%\keywords{Suggested keywords}%Use showkeys class option if keyword
                              %display desired
\maketitle

%\section{\label{sec:level1}First-level heading:\protect\\ The line
%break was forced \lowercase{via} \textbackslash\textbackslash}

One of the predictions shared by various quantum theories of gravity
is the existence of a minimum measurable distance, proportional to
the Planck length $\ell_p \sim (10^{-33}$cm) \cite{GUP source}, at
the Planck scale. This distance is caused by the nonlinear
interactions between gravity and matter. Uncertainty in the momentum
of a particle induces uncertainty on the geometry which leads to an
extra uncertainty to the position of the particle. Equivalently
speaking, there exists a cutoff on the energy scale, serving as an
ultimate ultraviolet regulator. One can refer to \cite{Garay:1994en}
for a review of the origins of the minimum measurable distance from
various scenarios.
%However, in the usual
%quantum mechanics, no constraint is imposed on the lower limit of
%detectable distance on the space.
In the spirit of the Hierarchy problem, one would naturally expect
that remnants of the minimum measurable distance phenomena show up
in effective theories at an up-to-determined intermediate scale.
%There are two possibilities at intermediate scales. Either the
%minimum measurable distance is suppressed by other gravity effects
%or prominent to reflect the influence of the gravity.
In this letter, when modeled with an extra dimension, we show that
the minimum measurable distance is the same scale as the radius of
the compactified extra dimension.

Some realizations of the minimum measurable distance are proposed.
One of the most important models is the generalized uncertainty
principle (GUP), derived from the modified fundamental
commutator\cite{Kempf,Brau,reviews,Das:2008kaa}:
\begin{equation}
[x,p]=if(p),
\end{equation}
where $f(p)$ is a positive function with $f(0)=1$ \footnote{In
\cite{Ali:2009zq}, $f(p)$ was assumed to negative possible.} to
reproduce the usual  fundamental commutator at the low energy limit.
We set $\hbar=c=1$ in this letter. The Taylor expansion of $f(p)$
around $p=0$ is:
\begin{equation}
f(p)=1+ \beta p^2 + {\cal{O}} (p^4),
\end{equation}
with $\beta=\beta_0 \ell_p^2 = \beta_0/M_p^2$ and $\beta_0$ is a
dimensionless number. $M_p\sim 10^{19}$GeV is the Planck mass. When
restricted to low energy effective theories, only the linear term of
$\beta$ is kept:
\begin{equation}
[x,p]=i(1+\beta p^2). \label{eq:1-D Comm}
\end{equation}
With this generalization, one can easily derive the GUP:
\begin{equation}
\Delta x \Delta p\geq \frac{1}{2}\big[1+\beta (\Delta p)^2\big],
\label{eq:1-DGUP}
\end{equation}
which in turn gives the minimum measurable distance $\Delta x \geq
\Delta x_{\min} =\sqrt\beta = \sqrt{\beta_0} \ell_p$. The tensorial
generalization of (\ref{eq:1-D Comm}) to higher dimensions is
represented by \footnote{The generic expression is $[x_i,
p_j]=i(\delta_{ij} +\beta \delta_{ij} p^2 + \beta' p_i p_j)$. We
choose $\beta'=2\beta$.}
\begin{equation}
[x_i, p_j]=i(\delta_{ij} +\beta \delta_{ij} p^2 + 2\beta p_i p_j),
\label{eq:GenComm}
\end{equation}
accompanied with $[x_i,x_j]=[p_i,p_j]=0$.
% to ensure the Jacobi identity.
Then the multidimensional version of GUP is
\begin{equation}
\Delta x_i \Delta p_i \geq \frac{1}{2} \Big[ 1+\beta\big((\Delta
p)^2 +\langle p\rangle ^2 \big) + 2\beta \big( \Delta p_i^2 +\langle
p_i\rangle ^2\big) \Big],
\end{equation}
where $i=1,2,3$ and $p^2 =\sum_{i=1}^3 p_i p_i$. The minimum
observable length for every direction is $\Delta (x_i)_{\min} \sim
\sqrt{\beta_0} \ell_p$. It should be emphasized that GUP is an low
energy approximation, capturing merely one of the features of the
physics at Planck scale. There exists no substantial reason to
expect that parameters in GUP are equal to those in full quantum
theories of gravity.

The dimensionless number $\beta_0=\beta/\ell_p^2 =\beta M_p^2$ plays
an important role. From eqn. (\ref{eq:1-DGUP}), in the GUP model, it
defines an intermediate energy scale where the gravity goes on
stage,
\begin{equation}
M_I\sim M_p/\sqrt\beta_0. \label{eq:EScale}
\end{equation}
Normally, $\beta_0$ is assumed to be of order unity. This makes
sense near the Planck scale, as indicated by various quantum
theories of gravity. However, when constructing low energy effective
theories, GUP for instance, this assumption has no firm basis. If
$\beta_0 \sim 1$, one obtains $M_I\sim M_{p}$. Nevertheless, once
approaching the Planck scale, the effective theory loses its
effectiveness and should be replaced by full quantum theories of
gravity. On the other hand, people tend to believe that there exist
some new physics at an intermediate scale $M_I$ $(<M_p)$, retaining
some features of the physics at the Planck scale. It is tempting to
conjecture that the minimum measurable length emerges already at
some $M_I$. At this intermediate scale, $\beta_0$ is not order of
unity anymore but a large number determined by eqn.
(\ref{eq:EScale}). The change of $\beta_0$ from order of unity to a
large number may be caused by some other suppressed features of the
full quantum theories of gravity.

The variation of $\beta_0$ could be justified by another
consideration. In physics, the existence of invariant dimensionless
parameters spoils the uniqueness of a theory. An immediate question
is that why our universe picks up a set of particular numbers. The
presence of dimensionless parameters either indicates that the
theory is an effective one or the parameters are not really
invariant with energy scales. Since $\beta_0$ originates from a full
quantum gravity theory, which is believed to be a final, unique
theory, it is hard to think that $\beta_0$ keeps fixed in all
scales.

Therefore, we conjecture that $\beta_0$ runs with the energy scale.
The story is not odd to us in quantum field theories (QFT). From the
renormalization group point of view, a cutoff $M_I$ of an effective
theory defines a grainy space with spacing $\Delta x =
f(\frac{1}{M_I})$, where $f$ is certain polynomial function. This
$\Delta x$ is believed to be caused by quantum fluctuation of
gravity and then could be an realization of the fundamental minimum
measurable length in a QFT. For this reason, the upper limit of
$\beta_0$ may exceed that set by the weak scale: $M_{\rm EW} \sim
10^2 $GeV. However, $\beta_0$ cannot increase forever to conflict
with experimental results in the usual quantum mechanics regime.

Consequently, from the weak scale, one obtains the roughest
estimation of the upper bound $\beta_0 < 10^{34}$. A better result
is presented by the conjectured genuine fundamental scale of ADD
model \cite{ArkaniHamed:1998rs} $M_I = M_f = 10^4$Gev, where the
effects of gravity are assumed to be easily visible, conjectured in
that model. With this scale,
\begin{equation}
\beta_0 \lesssim (M_p/M_f)^2 = 10^{30}.
\end{equation}
This upper bound is expected to be verified by LHC in the near
future. An up to date estimation $ \beta_0 < 10^{21}$ is given in
\cite{Das:2008kaa} with some lax assumptions. In their calculation
of some examples, the authors also showed that GUP effect is
unobservable with $\beta_0\sim 1$, consistent with our arguments
given above. This estimation indicates that the gravity becomes
important at $M_I\sim 10^9$GeV, much larger than the ADD scale.
Though out of the scope of terrestrial experiments, it is indeed a
reasonable intermediate scale. In this letter, we take an upper
bound based on the precision measurement of Lamb shift, given in
\cite{Das:2008kaa},
\begin{equation}
\beta_0 < 10^{36}.
\end{equation}
The corresponding scale is $M_I \sim 10$GeV. Joyfully it is in the
scope of current experiments. More accurate measurement is
anticipated in the near future to visualize the predictions of GUP
or to bring down the upper bound.
%However, as we explained above, the GUP is
%entangled with other effects, some delicately designed systems are
%in need to single out these solely from GUP. Furthermore,
%theoretical analysis is of great help.

On the other hand, though $\beta_0\sim 1$ near the Planck scale, in
a specific model, a lower bound on $\beta_0$ greater than unity
could exist due to the application range of the model. In the case
of GUP, a nice lower limit arises from the Grand Unification Theory
(GUT) scale in the Minimal Supersymmetric Standard Model (MSSM),
where the gravity becomes strong
\begin{equation}
M_I < M_{\rm GUT} \sim 10^{-3}M_{p}\Rightarrow \beta_0 > 10^6.
\end{equation}

At present, the existence of extra dimensions is widely accepted by
theorists. The story can be traced back to 1920's by the work of
Kaluza-Klein (KK). String theory demands ten dimensional spacetime
to have the anomalies canceled. Since the mid-1990's, this subject
has attracted large number of works. Several paradigms
\cite{Antoniadis:1990ew,
ArkaniHamed:1998rs,Randall:1999ee,Dvali:2000hr,Appelquist:2000nn,Cremades:2002dh,
Kokorelis:2002qi} are proposed and significant progress is achieved.
A good review is \cite{Shifman:2009df} and references therein.

Both minimum measurable distance and extra dimension are suggested
by quantum theories of gravity. It is therefore instructive to
combine them together in one effective theory. This reflection may
shed light on some information about $\beta_0$ or the scales of the
extra dimensions.

Guided with this idea, in this letter, based on GUP, we investigate
quantum systems with one extra dimension $w$ compactified on $S^1$
of radius $\rho$. We choose the KK compactification in this letter.
Alternative constructions based on minimum measurable length and
large extra dimensions are discussed in \cite{Hossenfelder:2003jz}
with various applications. Discussion on holographic counting
problems within the framework of extra dimensions and GUP can be
found in \cite{Scardigli:2003kr} with references. To implement the
generalized commutators (\ref{eq:GenComm}), one defines
\begin{eqnarray}
 &  & x_{i}=x_{0i},\nonumber \\
 &  & p_{i}=p_{0i}\left(1+\beta p_{0}^{2}\right),\quad
 p_{0i}=-i\frac{d}{dx_{0i}},
 \label{eq:iden}
 \end{eqnarray}
where $p_{0}^{2}=\sum p_{0j}p_{0j}$ and
$\left[x_{0i},p_{0j}\right]=i\delta_{ij}$, the usual canonical
operators. One can easily show that to the first order of $\beta$,
(\ref{eq:GenComm}) is guaranteed. For a quantum system described by
\begin{equation}
H=\frac{p^2}{2m} +V(\vec x),
\end{equation}
the modifications in (\ref{eq:iden}) can be treated as a
perturbation:
\begin{equation}
H= H_0 + H_1 =\frac{p_0^2}{2m} +V(\vec x_0) + \frac{\beta}{m} p_0^4.
\label{eq:PerturbH}
\end{equation}
The first order correction is then given as
\begin{eqnarray}
E_{n}^{\left(1\right)}&=&\frac{\beta}{m}\left\langle
n\right|p_{0}^{4}\left|n\right\rangle =4m\beta\left\langle
\left(E_n^{\left(0\right)}-V\right)^{2}\right\rangle,\nonumber\\
&=&4m\beta\left[\left(E_n^{\left(0\right)}\right)^{2}-
2E_n^{\left(0\right)} \left\langle V\right\rangle +\left\langle
V^{2}\right\rangle \right]. \label{eq:first order correction 2}
\end{eqnarray}

%In the ADD and Randall-Sundrum models, all particles carrying
%standard model charges (excitations of open strings) are confined to
%live on a $D3$ brane. Only graviton (excitations of closed strings)
%can propagate over the bulk space. There may exist several extra
%dimensions curled up with radii $\rho$. At first glance, it seems
%that our model is incompatible with these scenarios. However, we are
%considering a low energy effective quantum theory influenced by
%gravity. One is assured that some features of LED are demonstrated
%somehow. Furthermore, in the universal extra dimension (UED)
%scenario, all the matter fields in the standard model are free to
%move in the whole bulk space, though no compactification is assumed.
%As a heuristic argument, one is optional to conjecture that the $D3$
%brane, where open strings attached, is free to move along the
%compactified direction.

We discuss the simplest scenario in this letter. We suppose that in
the extra dimension, particles experience vanishing potential with
periodic boundary conditions. In a followed paper, we consider a
simple harmonic oscillator with non-zero potential in the extra
dimensions \cite{HO}. Thus, in the unperturbed system $H_0$ in
(\ref{eq:PerturbH}), the extra dimension contributes a term
\begin{equation}
\frac{\ell^2}{2m\rho^2},\quad\quad \ell=0,1,2\dots
\end{equation}
Therefore, the unperturbed eigenenergies in eqn. (\ref{eq:first
order correction 2}) should be replaced as
\begin{equation}
E_{n}^{\left(0\right)}\rightarrow {E_{n\ell}^{\left(0\right)}}'
\equiv E_{n}^{\left(0\right)} + \frac{\ell^2}{2m\rho^2},
\label{eq:ModE0}
\end{equation}
to reflect the modification from the extra dimension. The total
energy is then
\begin{equation}
E_{n\ell} = {E_{n\ell}^{\left(0\right)}}' +
4m\beta\left[\left({E_{n\ell}^{\left(0\right)}}'\right)^{2}-
2{E_{n\ell}^{\left(0\right)}}' \left\langle V\right\rangle
+\left\langle V^{2}\right\rangle \right]. \label{eq:totalE}
\end{equation}
Generically, $\langle V\rangle \simeq E_n^{\left(0\right)}$.
Therefore, when analyzing the order behaviors, on the {\it rhs} of
eqn. (\ref{eq:totalE}), the last two terms involving $\langle
V\rangle$ and $\langle V^2\rangle$ in the bracket can be ignored.
The spectra are approximated as
\begin{equation}
E_{n\ell} \simeq E_{n}^{\left(0\right)} \Big( 1+ 4m\beta
E_{n}^{\left(0\right)}\Big) + \Big( 1+ 8m\beta
E_{n}^{\left(0\right)} + 2\beta\frac{\ell^2}{\rho^2}\Big)
\frac{\ell^2}{2 m \rho^2}. \label{eq:ModTotalE}
\end{equation}
Consistency with the observations in quantum mechanics imposes
constraints on $\beta_0$ and $\rho$ as follows:
\begin{equation}
4m\beta E_{1}^{\left(0\right)}\ll 1\quad{\rm and}\quad \frac{1}{2 m
\rho^2} \gg E_1^{\left(0\right)}, \label{eq:ConsCond}
\end{equation}
or equivalently $\beta_0  \ll 10^{49}$ and $\rho \ll 10^{-9} {\rm
cm}$ with the \emph{ground state} scale $m E_1^{\left(0\right)}
\simeq 10^{-11} {\rm GeV}^2$ of typical quantum mechanics systems.
%For typical quantum systems with particle mass $m$, infinite square
%well or hydrogen atom, for instance, the energy scale is
%\begin{equation}
%m E_1^{\left(0\right)} \simeq 10^{-11} {\rm GeV}^2.
%\end{equation}
%In the first parenthesis on the {\it rhs} of eqn.
%(\ref{eq:ModTotalE}), consistency with the observations in quantum
%mechanics requires
%\begin{equation}
%4m\beta E_{1}^{\left(0\right)}\ll 1 \Rightarrow \beta_0 = \beta
%M_p^2 \ll 10^{50},
%\end{equation}
%consistent with the previous arguments. It should be emphasized
%that there must be
%\begin{equation}
%\frac{1}{2 m \rho^2} \gg E_1^{\left(0\right)} \Rightarrow \rho \ll
%10^{-8} {\rm cm},
%\end{equation}
%since no effect of gravity were observed at the scale of quantum
%mechanics.
From the spectra eqn. (\ref{eq:ModTotalE}), it is easy to see that
the gravity will not stage until $E_{n0}\sim E_{11}$, requiring
$E_{n}^{\left(0\right)}\sim\frac{1}{2 m \rho^2}$. To make this
condition possible, unbound up system is indicated in our model. We
are going to see that this is a very large scale compared with the
ground state one. It should be pointed out that the parameter $m$
here is the \emph{physical} mass determined by the scale, different
from the one in ground state. Therefore, the scale where gravity
becomes important is
\begin{equation}
E_{n0}\sim \frac{1}{2m\rho^2}\left(1+ 2\beta_0
\frac{\ell_p^2}{\rho^2} \right).
\end{equation}
The nice thing is that from the general analysis at eqn.
(\ref{eq:EScale}), the scale triggering gravity is $M_I=
M_p/\sqrt{\beta_0}= 1/\sqrt \beta_0\ell_p$. Thus, we have
\begin{equation}
\frac{1}{2M_I\rho^2}\left(1+ \frac{2}{M_I^2 \rho^2} \right) \sim
M_I,
\end{equation}
where $m$ has been replaced by the physical effective mass $M_I$.
With some simple algebraic calculation, it is easy to show that
%\begin{equation}
$\frac{1}{2M_I^2\rho^2} +  \Big(\frac{1}{M_I^2 \rho^2} \Big)^2 \sim
1$. Therefore, one simple and inspiring relation arises:
%\end{equation}
\begin{equation}
M_I \rho\sim 1 \quad{\rm or} \quad \beta = \beta_0\ell_p^2\sim
\rho^2. \label{eq:relation}
\end{equation}
Before discussing the physical significance of this relation, let us
loose the effectiveness of our model to the extreme situations
without much rigor. At the high energy end, Planck scale, T-duality
sets the minimum of the compatification radius $\rho_{\rm min} \sim
\sqrt{\alpha'} \sim \ell_p$, where $\alpha'$ is the regge slop. Then
eqn. (\ref{eq:relation}) gives $\beta_0\sim 1$, agreeing with the
prediction from various quantum gravity theories! In the low energy
limit, no evidence of extra dimension is ever observed. Effectively,
this means that $\rho$ approaches zero. It is then happy to find
that $\beta_0\to 0$, consistent with the fact that space is
continuously measurable. Though equation (\ref{eq:relation}) is
derived from an effective model, we showed that it also possesses
features of the extreme conditions.

%There are now three possibilities
%\begin{equation}
%\beta_0 \frac{\ell_p^2}{\rho^2} \gg 1, \quad \beta_0
%\frac{\ell_p^2}{\rho^2} \ll 1, \quad \beta_0
%\frac{\ell_p^2}{\rho^2} \sim 1.
%\end{equation}
%It is very nice that we have at least one boundary condition:
%$\beta_0 \sim 1$ at the Planck scale where $\rho \sim
%\sqrt{\alpha'} \sim \ell_p$ is predicted by T-duality. Therefore,
%the most probable relation is
%\begin{equation}
%\beta_0 \frac{\ell_p^2}{\rho^2} \sim 1.
%\end{equation}
Bearing in mind that there is an upper bound $\beta_0 < 10^{36}$
given by experimental results, one gets $\rho < 10^{-15}$cm. Though
this length is much smaller than the conjectured upper limit ($\sim
10^{-2}$cm) in ADD model, no conflict is found with any experiment
verified theory. The lower limit of the radius is given by $\rho >
10^{-30}$cm via the GUT scale of MSSM as we argued. We group the
results as follows:
%\begin{eqnarray}
%\beta_0 < 10^{34} &\Leftrightarrow & \rho < 10^{-16}{\rm cm}\\
%\rho \gg 10^{-22}{\rm cm} &\Leftrightarrow & \beta_0 \gg 10^{22},
%\end{eqnarray}
%giving the range of application of GUP model
\begin{equation}
10^{6} < \beta_0 < 10^{36},\quad 10^{-30}{\rm cm} < \rho <
10^{-15}{\rm cm},
\end{equation}
along with the scale $10{\rm GeV} < M_I < M_{\rm GUT}$. Hopfully,
the ranges could be narrowed by better refined models.

It is amazing that from eqn. (\ref{eq:relation}), one immediately
finds that the minimum measurable distance is exactly the radius of
the compact direction:
\begin{equation}
\Delta x_{\rm min} =\sqrt\beta=\sqrt\beta_0 \ell_p\sim \rho.
\end{equation}
Though it may look astonishing at first sight, one should not be
really surprised by this coincidence. The minimum measurable length
provides a cutoff to an effective theory while the compactification
radius defines the scale of the theory. Our results imply that once
an effective theory is constructed, its application range arises
simultaneously and the upper bound is close to its defining scale. A
more interesting implication of our derivations is that the
existence of minimum measurable length is probably nothing but the
exhibition of extra dimensions. It is still unclear if extra
dimensions can be detected by usual instruments other than
gravity-based devices. To our best knowledge, on the other hand, the
minimum measurable length is unreachable by apparatus based on gauge
particles.

To summarize, we argued that in GUP, an effective model, the
dimensionless parameter $\beta_0$ runs with energy scales. After
taking into account one compactified extra dimension, we showed that
the parameter $\beta =\beta_0 \ell_p^2$ is the same order of and
proportional to the compactification radius $\rho$. The predictions
are also consistent with results at Planck scale and usual quantum
mechanics scale. Rough ranges of $\beta_0$, $\rho$ and GUP scale
were also presented. Finally, we showed that the minimum measurable
length is precisely the compactification radius of the extra
dimension. We employed the simplest model in this letter. We hope
that better refined models based on our construction will reveal
more low energy consequences of quantum gravity and offer more
precise predictions, application range of GUP, for instance, in the
near future. Extensions include more compactified extra dimensions
corresponding different $\beta$'s, other compactification paradigms
like ADD or Sundrum-Randall models, introducing nonvanishing
periodic potentials on the extra dimensions and so on. More thorough
and detailed discussions on the running property of $\beta_0$ is
anticipated. Determining the upper limit of $\beta_0$ is of
particular interest. Probably, generalizing our construction to
quantum field theory is of help.

\begin{acknowledgments}
We are grateful to Bo Feng, Peng Wang and Qing-Hai Wang for
instructive discussions. We would also like to acknowledge useful
conversations with X. Guo and D. Chen. This work is supported in
part by NSFC (Grant No.10705008) and NCET.
\end{acknowledgments}


\begin{thebibliography}{99}

\small
\bibitem{GUP source}
G.~Veneziano,
%``A Stringy Nature Needs Just Two Constants,''
Europhys.\ Lett.\  {\bf 2}, 199 (1986);
%%CITATION = EULEE,2,199;%%
D.~J.~Gross and P.~F.~Mende,
%``String Theory Beyond the Planck Scale,''
Nucl.\ Phys.\  B {\bf 303}, 407 (1988);
%%CITATION = NUPHA,B303,407;%%
D.~Amati, M.~Ciafaloni and G.~Veneziano,
%``Can Space-Time Be Probed Below The String Size?,''
Phys.\ Lett.\  B {\bf 216}, 41 (1989);
%%CITATION = PHLTA,B216,41;%%
K.~Konishi, G.~Paffuti and P.~Provero,
%``Minimum Physical Length and the Generalized Uncertainty Principle
%in String Theory,''
Phys.\ Lett.\  B {\bf 234}, 276 (1990);
%%CITATION = PHLTA,B234,276;%%
R.~Guida, K.~Konishi and P.~Provero,
  %``On The Short Distance Behavior Of String Theories,''
Mod.\ Phys.\ Lett.\  A {\bf 6}, 1487 (1991).
%%CITATION = MPLAE,A6,1487;%%
M.~Maggiore,
%``A Generalized uncertainty principle in quantum gravity,''
Phys.\ Lett.\  B {\bf 304}, 65 (1993) [arXiv: hep-th/9301067].
%%CITATION = PHLTA,B304,65;%%

%\cite{Garay:1994en}
\bibitem{Garay:1994en}
  L.~J.~Garay,
  %``Quantum gravity and minimum length,''
  Int.\ J.\ Mod.\ Phys.\  A {\bf 10}, 145 (1995)
  [arXiv:gr-qc/9403008];
  %%CITATION = IMPAE,A10,145;%%
%
%\cite{Scardigli:1999jh}
%\bibitem{Scardigli:1999jh}
  F.~Scardigli,
  %``Generalized uncertainty principle in quantum gravity from micro-black  hole
  %gedanken experiment,''
  Phys.\ Lett.\  B {\bf 452}, 39 (1999)
  [arXiv:hep-th/9904025].
  %%CITATION = PHLTA,B452,39;%%



%\cite{Kempf:1994su}
\bibitem{Kempf}
  A.~Kempf, G.~Mangano and R.~B.~Mann,
  %``Hilbert Space Representation Of The Minimal Length Uncertainty Relation,''
  Phys.\ Rev.\  D {\bf 52}, 1108 (1995)
  [arXiv:hep-th/9412167];
  %%CITATION = PHRVA,D52,1108;%%
%
%\cite{Kempf:1996fz}
%\bibitem{Kempf:1996fz}
  A.~Kempf,
  %``Nonpointlike Particles in Harmonic Oscillators,''
  J.\ Phys.\ A  {\bf 30}, 2093 (1997)
  [arXiv:hep-th/9604045].
  %%CITATION = JPAGB,A30,2093;%%


%\cite{Brau:1999uv}
\bibitem{Brau}
  F.~Brau,
  %``Minimal Length Uncertainty Relation and Hydrogen Atom,''
  J.\ Phys.\ A  {\bf 32}, 7691 (1999)
  [arXiv:quant-ph/9905033];
  %%CITATION = JPAGB,A32,7691;%%
%
%\cite{Brau:2006ca}
%\bibitem{Brau:2006ca}
  F.~Brau and F.~Buisseret,
  %``Minimal length uncertainty relation and gravitational quantum well,''
  Phys.\ Rev.\  D {\bf 74}, 036002 (2006)
  [arXiv:hep-th/0605183].
  %%CITATION = PHRVA,D74,036002;%%

\bibitem{reviews}
%\cite{Chang:2001kn}
%\bibitem{Chang:2001kn}
  L.~N.~Chang, D.~Minic, N.~Okamura and T.~Takeuchi,
  %``Exact solution of the harmonic oscillator in arbitrary dimensions with
  %minimal length uncertainty relations,''
  Phys.\ Rev.\  D {\bf 65}, 125027 (2002)
  [arXiv:hep-th/0111181v2];
  %%CITATION = PHRVA,D65,125027;%%
%
%\cite{Dadic:2002qn}
%\bibitem{Dadic:2002qn}
  I.~Dadic, L.~Jonke and S.~Meljanac,
  %``Harmonic oscillator with minimal length uncertainty relations and ladder
  %operators,''
  Phys.\ Rev.\  D {\bf 67}, 087701 (2003)
  [arXiv:hep-th/0210264];
  %%CITATION = PHRVA,D67,087701;%%
%
%\cite{Nozari:2005mr}
%\bibitem{Nozari:2005mr}
  K.~Nozari and T.~Azizi,
  %``Some Aspects of Minimal Length Quantum Mechanics,''
  Gen.\ Rel.\ Grav.\  {\bf 38}, 735 (2006)
  [arXiv:quant-ph/0507018];
  %%CITATION = GRGVA,38,735;%%
%
%\bibitem{S reduction 1}
  M. M. Stetsko, V. M. Tkachuk, e-print (2006),
  [arXiv:quant-ph/0603042v1];
%
%\bibitem{S reduction 2}
M. M. Stetsko, Phys. Rev. \textbf{A} \textbf{74}, 062105 (2006)
[arXiv:quant-ph/0703269v1];
%
%\cite{Benczik:2007we}
%\bibitem{Benczik:2007we}
  S.~Z.~Benczik,
  ``Investigations on the minimal-length uncertainty relation,''
  Dissertation for Doctor of Philosophy;
  %%CITATION = UMI-32-49454;%%
%
%\cite{Battisti:2007jd}
%\bibitem{Battisti:2007jd}
  M.~V.~Battisti and G.~Montani,
  %``The big-bang singularity in the framework of a generalized uncertainty
  %principle,''
  Phys.\ Lett.\  B {\bf 656}, 96 (2007)
  [arXiv:gr-qc/0703025];
  %%CITATION = PHLTA,B656,96;%%
%
%\cite{Battisti:2007zg}
%\bibitem{Battisti:2007zg}
  M.~V.~Battisti and G.~Montani,
  %``Quantum Dynamics of the Taub Universe in a Generalized Uncertainty
  %Principle framework,''
  Phys.\ Rev.\  D {\bf 77}, 023518 (2008)
  [arXiv:0707.2726 [gr-qc]];
  %%CITATION = PHRVA,D77,023518;%%
%
%\cite{Battisti:2008du}
%\bibitem{Battisti:2008du}
  M.~V.~Battisti,
  %``Loop and braneworlds cosmologies from a deformed Heisenberg algebra,''
  arXiv:0805.1178 [gr-qc].
  %%CITATION = ARXIV:0805.1178;%%

%\cite{Das:2008kaa}
\bibitem{Das:2008kaa}
  S.~Das and E.~C.~Vagenas,
  %``Universality of Quantum Gravity Corrections,''
  Phys.\ Rev.\ Lett.\  {\bf 101}, 221301 (2008)
  [arXiv:0810.5333 [hep-th]].
  %%CITATION = PRLTA,101,221301;%%

%\cite{Ali:2009zq}
\bibitem{Ali:2009zq}
  A.~F.~Ali, S.~Das and E.~C.~Vagenas,
  %``Discreteness of Space from the Generalized Uncertainty Principle,''
  Phys.\ Lett.\  B {\bf 678}, 497 (2009)
  [arXiv:0906.5396 [hep-th]].
  %%CITATION = PHLTA,B678,497;%%

%\cite{Antoniadis:1990ew}
\bibitem{Antoniadis:1990ew}
  I.~Antoniadis,
  %``A Possible new dimension at a few TeV,''
  Phys.\ Lett.\  B {\bf 246}, 377 (1990).
  %%CITATION = PHLTA,B246,377;%%

%\cite{ArkaniHamed:1998rs}
\bibitem{ArkaniHamed:1998rs}
  N.~Arkani-Hamed, S.~Dimopoulos and G.~R.~Dvali,
  %``The hierarchy problem and new dimensions at a millimeter,''
  Phys.\ Lett.\  B {\bf 429}, 263 (1998)
  [arXiv:hep-ph/9803315];
  %%CITATION = PHLTA,B429,263;%%
%\cite{Antoniadis:1998ig}
%\bibitem{Antoniadis:1998ig}
  I.~Antoniadis, N.~Arkani-Hamed, S.~Dimopoulos and G.~R.~Dvali,
  %``New dimensions at a millimeter to a Fermi and superstrings at a TeV,''
  Phys.\ Lett.\  B {\bf 436}, 257 (1998)
  [arXiv:hep-ph/9804398].
  %%CITATION = PHLTA,B436,257;%%


%\cite{Randall:1999ee}
\bibitem{Randall:1999ee}
  L.~Randall and R.~Sundrum,
  %``A large mass hierarchy from a small extra dimension,''
  Phys.\ Rev.\ Lett.\  {\bf 83}, 3370 (1999)
  [arXiv:hep-ph/9905221];
  %%CITATION = PRLTA,83,3370;%%
%
%\cite{Randall:1999vf}
%\bibitem{Randall:1999vf}
  L.~Randall and R.~Sundrum,
  %``An alternative to compactification,''
  Phys.\ Rev.\ Lett.\  {\bf 83}, 4690 (1999)
  [arXiv:hep-th/9906064].
  %%CITATION = PRLTA,83,4690;%%

%\cite{Dvali:2000hr}
\bibitem{Dvali:2000hr}
  G.~R.~Dvali, G.~Gabadadze and M.~Porrati,
  %``4D gravity on a brane in 5D Minkowski space,''
  Phys.\ Lett.\  B {\bf 485}, 208 (2000)
  [arXiv:hep-th/0005016].
  %%CITATION = PHLTA,B485,208;%%

%\cite{Appelquist:2000nn}
\bibitem{Appelquist:2000nn}
  T.~Appelquist, H.~C.~Cheng and B.~A.~Dobrescu,
  %``Bounds on universal extra dimensions,''
  Phys.\ Rev.\  D {\bf 64}, 035002 (2001)
  [arXiv:hep-ph/0012100].
  %%CITATION = PHRVA,D64,035002;%%

%\cite{Cremades:2002dh}
\bibitem{Cremades:2002dh}
  D.~Cremades, L.~E.~Ibanez and F.~Marchesano,
  %``Standard model at intersecting D5-branes: Lowering the string scale,''
  Nucl.\ Phys.\  B {\bf 643}, 93 (2002)
  [arXiv:hep-th/0205074];
  %%CITATION = NUPHA,B643,93;%%

%\cite{Kokorelis:2002qi}
\bibitem{Kokorelis:2002qi}
  C.~Kokorelis,
  %``Exact standard model structures from intersecting D5-branes,''
  Nucl.\ Phys.\  B {\bf 677}, 115 (2004)
  [arXiv:hep-th/0207234].
  %%CITATION = NUPHA,B677,115;%%


%\cite{Shifman:2009df}
\bibitem{Shifman:2009df}
  M.~Shifman,
  %``LARGE EXTRA DIMENSIONS: Becoming acquainted with an alternative paradigm,''
  arXiv:0907.3074 [hep-ph].
  %%CITATION = ARXIV:0907.3074;%%

%\cite{Hossenfelder:2003jz}
\bibitem{Hossenfelder:2003jz}
  S.~Hossenfelder, M.~Bleicher, S.~Hofmann, J.~Ruppert, S.~Scherer and H.~Stoecker,
  %``Collider signatures in the Planck regime,''
  Phys.\ Lett.\  B {\bf 575}, 85 (2003)
  [arXiv:hep-th/0305262];
  %%CITATION = PHLTA,B575,85;%%
%\cite{Hossenfelder:2004ze}
%\bibitem{Hossenfelder:2004ze}
  S.~Hossenfelder,
  %``Suppressed black hole production from minimal length,''
  Phys.\ Lett.\  B {\bf 598}, 92 (2004)
  [arXiv:hep-th/0404232];
  %%CITATION = PHLTA,B598,92;%%
%\cite{Hossenfelder:2004up}
%\bibitem{Hossenfelder:2004up}
  S.~Hossenfelder,
  %``Running coupling with minimal length,''
  Phys.\ Rev.\  D {\bf 70}, 105003 (2004)
  [arXiv:hep-ph/0405127];
  %%CITATION = PHRVA,D70,105003;%%
%\cite{Hossenfelder:2004wv}
%\bibitem{Hossenfelder:2004wv}
  S.~Hossenfelder,
  %``Large extra dimensions and the minimal length,''
  Czech.\ J.\ Phys.\  {\bf 55}, B809 (2005)
  [arXiv:hep-ph/0409350];
  %%CITATION = CZYPA,55,B809;%%
%\cite{Harbach:2005yu}
%\bibitem{Harbach:2005yu}
  U.~Harbach and S.~Hossenfelder,
  %``The Casimir effect in the presence of a minimal length,''
  Phys.\ Lett.\  B {\bf 632}, 379 (2006)
  [arXiv:hep-th/0502142].
  %%CITATION = PHLTA,B632,379;%%


%\cite{Scardigli:2003kr}
\bibitem{Scardigli:2003kr}
  F.~Scardigli and R.~Casadio,
  %``Generalized uncertainty principle, extra-dimensions and holography,''
  Class.\ Quant.\ Grav.\  {\bf 20}, 3915 (2003)
  [arXiv:hep-th/0307174];
  %%CITATION = CQGRD,20,3915;%%
%\cite{Scardigli:2007bw}
%\bibitem{Scardigli:2007bw}
  F.~Scardigli and R.~Casadio,
  %``Is the Equivalence Principle violated by Generalized Uncertainty Principles
  %and Holography in a brane-world?,''
  Int.\ J.\ Mod.\ Phys.\  D {\bf 18}, 319 (2009)
  [arXiv:0711.3661 [hep-th]].
  %%CITATION = IMPAE,D18,319;%%




\bibitem{HO}
P. ~Wang, H. ~Wu and H. ~Yang, ``Spectra of harmonic oscillators
with GUP and extra dimensions,'' to appear.


\end{thebibliography}
\end{document}